\documentclass[aps,amsfonts,amssymb,nofootinbib,twocolumn,superscriptaddress]{revtex4}
\usepackage{graphics}
\usepackage{hyperref}
\usepackage{epsfig}
\usepackage{color}

\begin{document}

\bibliographystyle{naturemag}

\title{A sudden collapse in the transport lifetime across the topological phase transition in (Bi$_{1-x}$In$_x$)$_2$Se$_3$}



\author{Liang Wu}
\affiliation{The Institute for Quantum Matter, Department of Physics and Astronomy, The Johns Hopkins University, Baltimore, MD 21218 USA.}
\author{M. Brahlek}
\affiliation{Department of Physics and Astronomy, Rutgers the State University of New Jersey. Piscataway, NJ 08854}
\author{R. Vald\'es Aguilar}
\affiliation{The Institute for Quantum Matter, Department of Physics and Astronomy, The Johns Hopkins University, Baltimore, MD 21218 USA.}
\author{A. V. Stier}
\affiliation{The Institute for Quantum Matter, Department of Physics and Astronomy, The Johns Hopkins University, Baltimore, MD 21218 USA.}
\author{C. M. Morris}
\affiliation{The Institute for Quantum Matter, Department of Physics and Astronomy, The Johns Hopkins University, Baltimore, MD 21218 USA.}
\author{Y. Lubashevsky}
\affiliation{The Institute for Quantum Matter, Department of Physics and Astronomy, The Johns Hopkins University, Baltimore, MD 21218 USA.}
\author{L. S. Bilbro}
\affiliation{The Institute for Quantum Matter, Department of Physics and Astronomy, The Johns Hopkins University, Baltimore, MD 21218 USA.}
\author{N. Bansal}
\affiliation{Department of Physics and Astronomy, Rutgers the State University of New Jersey. Piscataway, NJ 08854}
\author{S. Oh}
\affiliation{Department of Physics and Astronomy, Rutgers the State University of New Jersey. Piscataway, NJ 08854}
 \author{N. P. Armitage}
 \email{npa@pha.jhu.edu}
 \affiliation{The Institute for Quantum Matter, Department of Physics and Astronomy, The Johns Hopkins University, Baltimore, MD 21218 USA.}

 \date{\today}

\maketitle

\textbf{Topological insulators (TIs) are newly discovered states of matter with robust metallic surface states protected by the topological properties of the bulk wavefunctions \cite{Hasan-Kane-10,Hasan-Moore-10,Qi-Zhang-11,FuKaneMele07,MooreBalents07,Roy09a}.  A quantum phase transition (QPT) from a TI to a conventional insulator and a change in topological class can only occur when the bulk band gap closes \cite{Qi-Zhang-11}.  In this work, we have utilized time-domain terahertz spectroscopy  (TDTS) to investigate the low frequency conductance in (Bi$_{1-x}$In$_x$)$_2$Se$_3$ as we tune through this transition by indium substitution.  Above certain substitution levels we observe a collapse in the transport lifetime that indicates the destruction of the topological phase.  We associate this effect with the threshold where states from opposite surfaces hybridize.   The substitution level of the threshold is thickness dependent and only asymptotically approaches the bulk limit $x \approx 0.06$ where a maximum in the mid-infrared absorption is exhibited.  This absorption can be identified with the bulk band gap closing and a change in topological class.   The correlation length associated with the QPT appears as the evanescent length of the surface states.  The observation of the thickness-dependent collapse of the transport lifetime shows the unusual role that finite size effects play in this topological QPT. }

\bigskip

\begin{figure*}
\begin{center}
\includegraphics[width=1\textwidth]{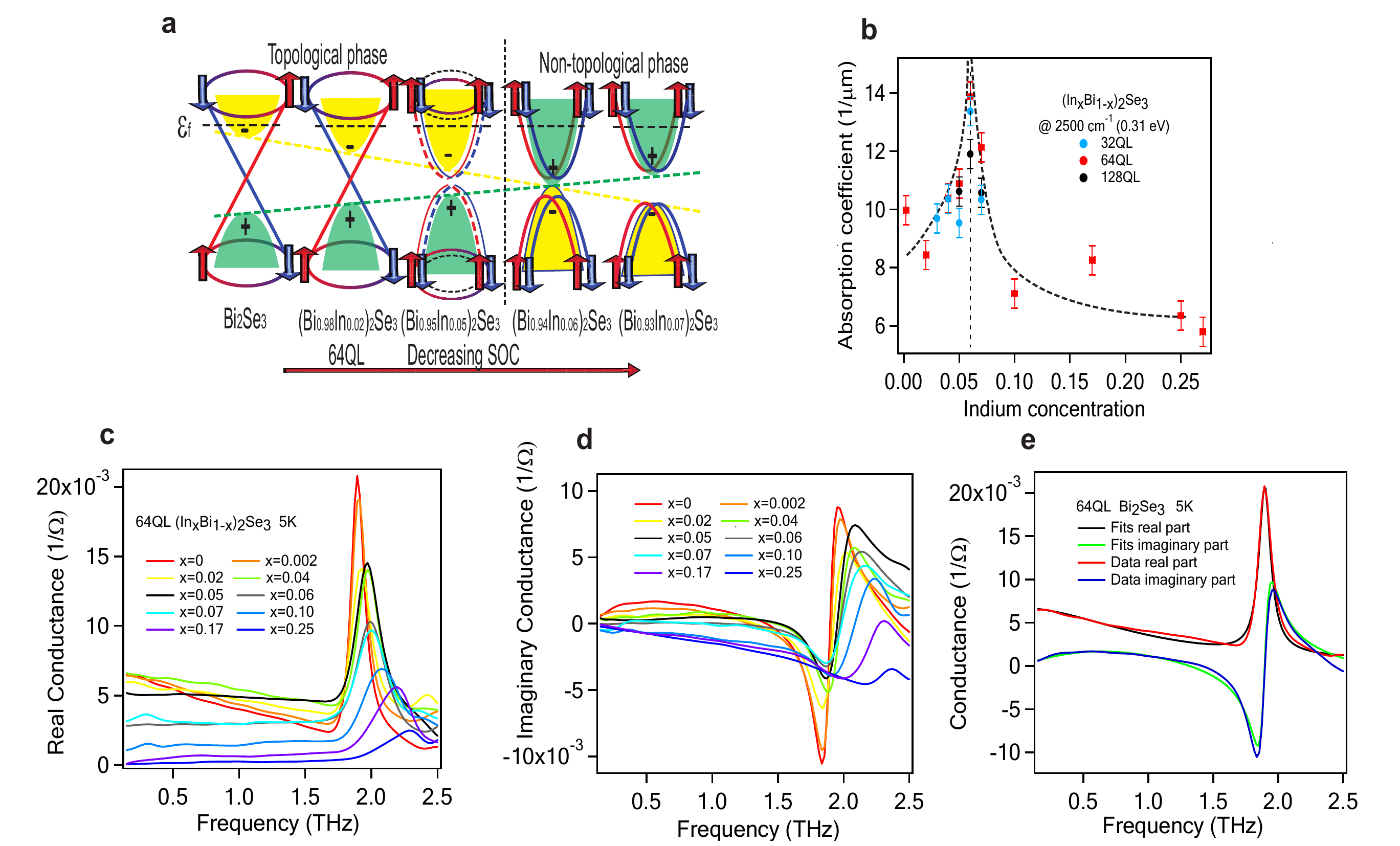}
\centering
\caption{ \textsf{\textbf{MIR absorption and THz sheet conductance of (Bi$_{1-x}$In$_x$)$_2$Se$_3$ films.}}
\textsf{\textbf{a,}} A schematic of bulk band inversion for the 64 QL series.  Valence and conduction band parity is indicated by the solid yellow-green color and the plus-minus sign.  Spin branches in the non-TI regime are shifted for clarity.  \textsf{\textbf{b,}} Room temperature mid-infrared absorption ($\alpha=ln(1/T)/d$, $T$ is transmission and $d$ is film thickness) at 0.31 eV (2500 cm$^{-1}$) as a function of indium concentration for 32 QL, 64 QL and 128 QL. The error bar derives from the thickness variation in substrates. The curved dashed line is a guide to the eye. The vertical dash line indicates where the bulk band gap closes.    \textsf{\textbf{c,}}  Real and \textsf{\textbf{d,}}  Imaginary parts of the THz sheet conductance  of 64 QL (Bi$_{1-x}$In$_x$)$_2$Se$_3$  films for different indium levels at 5K.  \textsf{\textbf{e,}}  Experimental data with fits for 64 QL Bi$_2$Se$_3$ at 5K.} 
\label{Conductance}
\end{center}
\end{figure*}

The topological character  of TIs is determined by the nature of their valence-band wave functions, which can be quantified by 4 $Z_{2}$ invariants.  Fu and Kane have shown that for inversion symmetric crystals it is possible to evaluate these invariants directly with knowledge of the parity of Bloch wave functions for the occupied electronic states at high symmetry points in the Brillouin zone\cite{Fu07a}.   Although their argument is formulated for inversion symmetric systems, a material's topological classification does not require inversion or translation symmetry. Therefore the expectation is that the alloying of known TIs with lighter elements by reducing spin-orbit coupling or the tuning of lattice constant can cause the bulk band gap $\Delta$ to close and invert at a quantum critical point where the topological class changes (See cartoon Fig. \ref{Conductance}\textbf{a}).  This has been investigated in the thallium-based ternary chalcogenide alloy TlBi(S$_{1-x}$Se$_x$)$_2$ \cite{Xu11,Sato11,Souma12}, but thus far only with photoemission (Supplementary Information (SI) section B).  Although signatures of topological surface state (TSS) conduction have been found in Bi$_2$Se$_3$ \cite{Qu10a,Taskin12,Rolando_Kerr,Bansal12a}, a demonstration that the surface transport changes dramatically when the band gap closes and the bulk changes topological class \cite{Goswami11a} would be strong evidence for the topological nature of these materials and is still lacking.    In this regard, it was pointed out recently that indium (In) substitutes for bismuth to form a solid solution in Bi$_2$Se$_3$ and that the non-topological end member In$_2$Se$_3$ of the (Bi$_{1-x}$In$_x$)$_2$Se$_3$ series shares the common rhombohedral  D$^{5}_{3d}$ structure with Bi$_2$Se$_3$\cite{Brahlek12}.  In Ref. \cite{Brahlek12} a topological to trivial transition was observed in a range $x\sim 0.03-0.07$. In this work, we systematically investigate the details of this topological phase transition.

Samples investigated in this work were films grown by molecular beam epitaxy.   Bi$_2$Se$_3$ films grow in ``quintuple layer" (QL) Se-Bi-Se-Bi-Se blocks.  In addition to TDTS, we performed optical absorption measurements in the mid-infrared range, which are sensitive to the energy scale of the bulk band gap ($\sim$350 meV for Bi$_2$Se$_3$).  In Fig. \ref{Conductance}\textbf{b} we show that the MIR absorption coefficient at  0.31 eV has a distinct peak near $x=0.06$ for all the 32, 64, and 128 QL thicknesses.  Despite the fact that even the $x=0$ compound has its chemical potential in the bottom of the conduction band, a closing of the band gap should increase the absorption as the low energy joint density of states increases.  As we will show below, the peak in the MIR absorption with In substitution is consistent with a topological phase transition occurring by closing and reopening of the band gap at $x \approx 0.06$. 

In Fig. \ref{Conductance}\textbf{c,d} we show the real and imaginary THz range (1 THz = 4.1 meV) conductance of a series of 64 QL (Bi$_{1-x}$In$_x$)$_2$Se$_3$ films substituted from $x=0$ to $x=0.25$ at 5K in the THz frequency range.    For $x=0$ we have previously shown (and will demonstrate again below) that the spectra are characterized by a prominent low frequency Drude component, which has an almost entirely surface character, and a low frequency phonon at $\sim$1.9 THz\cite{Rolando_Kerr}.   With increasing In concentration, the phonon mode broadens and its frequency shifts almost linearly (SI Fig. 4\textbf{e}) to higher frequency due to the lighter indium mass.  The Drude peak broadens only slightly up to $x=0.04$ in a fashion expected due to the increased disorder.  However for substitutions larger than $x=0.05$ the Drude peak broadens dramatically and the real part of the conductance becomes flat, indicating that the TSSs have undergone a significant change\cite{Brahlek12}.  Qualitatively similar results at different doping levels are seen for 16 QL, 32 QL and 128 QL films (SI section C). For $x\geq0.10$,  the conductance becomes a slowly increasing function of frequency $\omega$ consistent with a tendency towards charge localization.  The finite low frequency conductance for $x=0.10$ indicates a reminiscent of metallic behavior, possibly contributed by the electron pocket seen by photoemission at these substitutions \cite{Brahlek12}.  The nearly zero conductance of $x=0.17$ and $x=0.25$ at low frequency and correspondingly the negative slope of the imaginary conductance at low $\omega$ indicates insulating behavior for these concentration levels, which is consistent with the DC study\cite{Brahlek12}.

\begin{figure*}
\begin{center}
\includegraphics[width=1\textwidth]{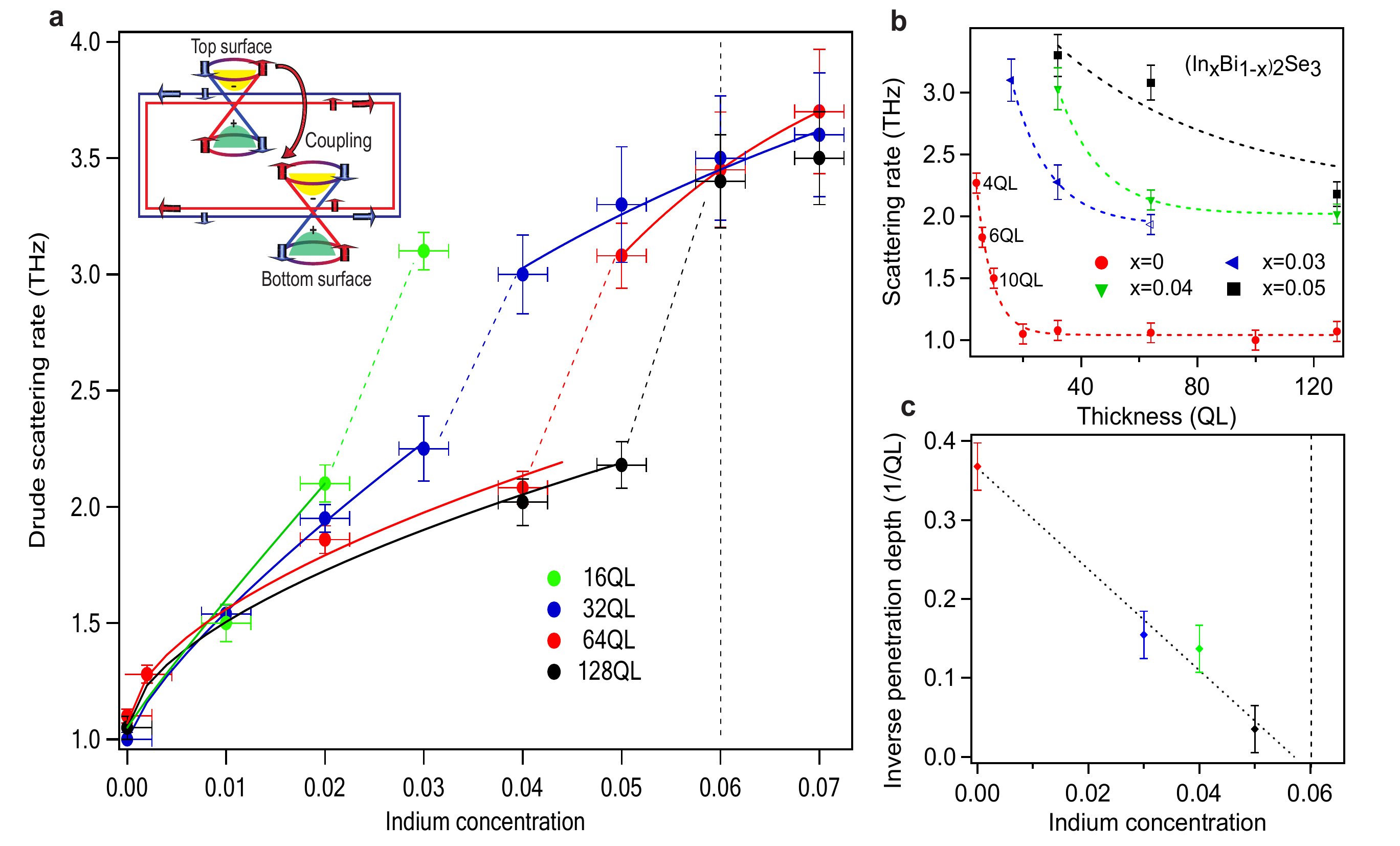}
\centering
\caption{\textsf{\textbf{Finite size effects and evidence for topological phase transition.} }\textsf{\textbf{a,}}  Scattering rate $1/\tau$ ($\tau$ is the transport lifetime) of the Drude term from fits as a function of indium substitution.  The error bars are defined as the parameter range where acceptable fits are obtained.  Lines are guides to the eye.  The vertical black dashed line indicates where bulk band gap closes and the bands invert. \textsf{\textbf{Inset:}} A schematic explaining the jump in scattering rate, in which back scattering can occur when surfaces are coupled. \textsf{\textbf{b,}}  The scattering rate as a function of thickness at different In concentrations. The dashed lines are fits $1/\tau= (A+B  e^{-d/2\xi})$. $\xi$ is the TSS penetration depth and d is the film thickness in QLs. The 64 QL $x=0.03$ point is an interpolation from `\textsf{\textbf{a}}'.  \textsf{\textbf{c,}}  Inverse penetration depth $1/\xi$ as a function of In concentration. The dashed line is a linear fit.  Error bars are the uncertainties in the fits in  `\textsf{\textbf{b}}'.}
\label{FitResults}
\end{center}
\end{figure*}

To investigate the details of the evolution through the topological QPT, we fit the data to a three component Drude-Lorentz model as shown in Fig. \ref{Conductance}\textbf{e} (See Methods and SI).   Excellent fits are obtained for all data for $x<0.08$.  For the Drude component these fits give a spectral width that can be associated with the transport scattering rate  $1/\tau$ and a spectral weight that can be shown to be a function of carrier density.  Concentrating again on the 64 QL sample, the Drude scattering rate increases smoothly for low substitutions (Fig. \ref{FitResults}\textbf{a}), which is consistent with our above observations and expectation due to increased impurity scattering.  However at $x=0.05$ the smooth trend suddenly changes and the scattering rate is enhanced dramatically.  Just by In substitution alone, we would expect a continued smooth increase of the scattering rate and therefore the sudden change at $x=0.05$ in the 64 QL film indicates a transition occurring in the TSSs. We can correlate the sudden changes in the surface transport properties with the MIR absorption peak which is associated with the closing of the bulk band gap.  We conclude that with In substitution a topological phase transition occurs by the closing and reopening of the band gap at $x \approx 0.06$.  Therefore, we associate the sudden collapse in the transport lifetime with a loss of topological protection of the surface states upon entering the non-topological state.  Note that the spectral weight of the Drude component also changes as a function of indium substitution (See SI Fig. 4\textbf{f}). The origin of the change is unclear, but may arise due to a shift of spectral weight from higher energies or a change in the surface state Fermi velocity.

In general one expects that the penetration length of the TSS wave function is proportional to $\hbar v_F/\Delta$ \cite{Linder09a,ZhangMele12}.  In the topological regime, $180^\circ$ backscattering is prohibited due to spin-momentum locking \cite{Fu07a}.   However, $180^\circ$ backscattering becomes possible when the the band gap closes sufficiently such that the wave function penetration is of order half the film thickness, TSSs can hybridize and an electron in a state $k$ with a particular spin can scatter into a state $-k$ with the same spin that derives from the other surface, as shown by the inset of Fig. \ref{FitResults}\textbf{a}. One expects that when topological protection is lost in this fashion, the phase space for scattering increases by a factor of order unity (SI section D), as observed.  Our results show directly how the robust transport properties and topological protection of the surface states are dependent on bulk band inversion, but also the essential role of finite size effects in such transitions.

\begin{figure*}
\begin{center}
\includegraphics[width=1\textwidth]{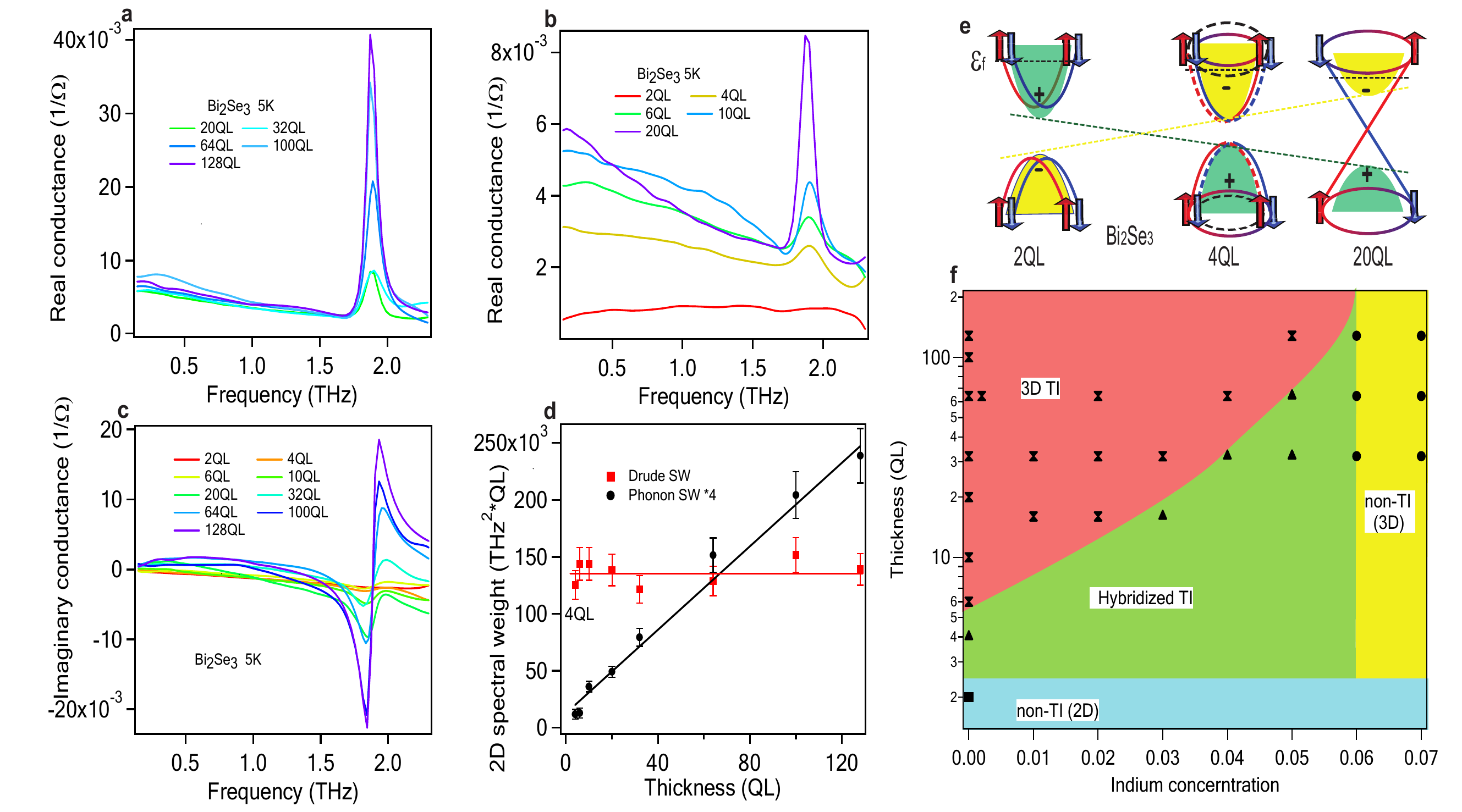}
\centering
\caption{\textsf{\textbf{Crossover from 3D TI to 2D TI in pure Bi$_2$Se$_3$.}} \textsf{\textbf{a, b,}} Real  and  \textsf{\textbf{c,}} imaginary sheet conductance of pure Bi$_2$Se$_3$ films with different thickness at 5K.  \textsf{\textbf{d,}} 2D Drude spectral weight  (($\omega_p/2\pi)^2 d $) and 2D phonon spectral weight ($4 (\omega_{p - phonon}/2\pi) ^2 d $ )  as a function of film thickness.  \textsf{\textbf{e,}} A schematic showing how TSS hybridization occurs when the film thickness is reduced. The 4 QL and 2 QL films may be topologically non-trivial and trivial respectively as proposed in \cite{Liu-ZhangPRB10, Lu-ShenPRB10}.  Spin-up and spin-down branches in the non-TI regime are shifted for clarity.   \textsf{\textbf{f,}} A phase diagram constructed to distinguish different phases. Solid markers with different shapes correspond to data in different regimes.   The boundary between TI and the hybridized regime is set by the length scales determined by the fits in Fig. \ref{FitResults} \textsf{\textbf{b}}.}

\label{Thickness}
\end{center}
\end{figure*}

Looking closer we note that in fact for the 64 QL sample we observe the threshold for an increased THz scattering rate at $x=0.05$, while the MIR absorption peaks at $x=0.06$.  If the surface state's evanescent length is of order a few nm at $x=0$,  a simple estimate based on linearly extrapolating its 0.35 eV band gap to close at $x=0.06$ shows that a 64 QL film should have the threshold for the two TSSs to hybridize and increase scattering shift downward by an amount of order $x=0.01$ (SI Fig. 3). Reducing the film thickness makes TSS hybridization occur at even lower In concentration (SI Fig. 3).  This dependence of the transport lifetime on film thickness can be seen directly by looking in detail at the fit results  in Fig. \ref{FitResults}\textbf{a} for 16, 32, and 128 QL series samples.  For the same In substitution, thinner samples generally show a more broadened Drude component (SI section C). When the film thickness is reduced, the jump in the scattering rate occurs at $x=0.03$ and $x=0.04$ for the 16 QL and 32 QL series respectively. In contrast, for thicker 128 QL samples, the jump in transport (at $x=0.06$) approaches the substitution level where the MIR peak occurs.  Therefore, we infer that $x \approx 0.06$ is the actual critical concentration where band inversion occurs but the jump in transport occurs at lower concentration in thin samples due to the finite size effect.

Related physics can be observed by performing thickness dependent studies. One does not induce a QPT $per$ $se$, but one can observe the increased phase space due to scattering as surface states overlap.  In Fig. \ref{Thickness}\textbf{a-c} we show the real and imaginary part of the sheet conductance of a pure Bi$_2$Se$_3$ films with different thickness. As is clearly shown in Fig. \ref{Thickness}\textbf{a, b} the low frequency Drude component is reasonably thickness independent above 20 QL and only broadens appreciably near 10 QL.  The Drude component broadens smoothly below 10 QL and can be resolved well down to 4 QL, while the 2 QL sample shows completely different behavior. We apply the Drude-Lorentz model again to investigate the detailed evolution of the thickness dependence. The 2 QL sample is excluded because no clear Drude peak is observed. As shown in Fig. \ref{Thickness}\textbf{d}, the Drude spectral weight (SW) is independent of thickness and the phonon spectral weight increases linearly with thickness, which agrees with our previous report\cite{Rolando_Kerr}.  The Drude scattering rate is enhanced dramatically for film thickness below 10 QL. As shown in Fig. \ref{FitResults}\textbf{b},  the dependence can be  fit well within a Fermi's golden rule approach by an exponential function that models the overlap between surfaces $1/\tau= (A+B e^{-d/2\xi})$ where $\xi = 2.7\pm0.3$ QL.   This evanescent length agrees well with estimates for the extent of the surface states at $x=0$ from ARPES measurements\cite{Zhang-Xue10}.  Changes in DC transport have also been found below these thicknesses and interpreted as the loss of topological protection\cite{Bansal12a,Taskin12}.  We note that the fact that we extract approximately the same evanescent decay length for the states at E$_F$ as for the states  near the Dirac point from ARPES is a particular feature of the topological surface band structure.   As long as one considers high symmetry surfaces the decay length can be shown to be independent of energy for all energies below the conduction band threshold \cite{ZhangMele12,ZhangMelePrivateComm} (SI section A for further information).

One can apply the scaling analysis as a function of thickness (Fig. \ref{FitResults}\textbf{b}) to extract the relevant length scale for the Indium substituted samples.   The advantage of performing this scaling as a function of thickness is that a sample's disorder level remains constant.   In each of these fits we constrain the constant offset $A$ to be the value of the scattering rate of the thickest sample.   Showing a remarkable consistency in our treatment, one can see from Fig. \ref{FitResults}\textbf{c} that the inverse length scale extrapolates linearly to zero at the substitution level of the QPT $x \approx 0.06$ in a fashion expected for a correlation length near a continuous transition.

It is interesting to note the dramatic difference in spectra between 4 QL and 2 QL $x=0$ samples. An oscillatory behavior between 2D topological nontrivial and trivial behavior as a function of thickness has been theoretically proposed\cite{Liu-ZhangPRB10, Lu-ShenPRB10}, as shown schematically in Fig. \ref{Thickness}\textbf{e}.  The large differences we observe could be related to such a phase transition.  Nevertheless, one must be careful here as surface steps with 1 QL depth and oxidization are very important in the 2 QL sample and, therefore,  it is impossible to make a strong conclusion (Additional discussion in SI section G).

 The large number of samples we have measured and the thickness dependent fits allow us to map out a detailed ``phase diagram" as function of thickness and concentration that includes regions of 3D TI, 3D non-TI, and a regime we call ``hybridized TI", where the bulk bands are still inverted but surface states hybridize.   As shown in Fig. \ref{Thickness}\textbf{f} the threshold for increased scattering starts near 5 QL (the point at which surface states penetrating $\sim 2.5 $ QL first overlap) and asymptotically approaches the boundary where the ``true" phase transition occurs in a thermodynamically large sample.  Note that increased scattering can be resolved even for films thicker than this threshold as the involved wavefunctions have an exponential tail.

In the non-topological phase an even number of spin species must exist on a surface as opposed to the topological phase where an odd number of spin species exist.   Our work demonstrates where the other spin species ``comes from" when a system crosses the QPT.  It comes from the ``other" surface.  Our work also shows the non-trivial effects that the finite size of a sample has on a QPT as a consequence of the bulk-boundary correspondence-derived properties of the TSS.  Here the role of the correlation length associated with the QPT is played by the evanescent decay length of the TSS.  It would be interesting to investigate these effects even asymptotically closer to the transition.  We also hope this work generates further interest in the field of topological phase transitions, particularly as a guide for first principle calculations for In substitution in Bi$_2$Se$_3$ to reveal the evolution of the band structure. Our work will also facilitate applications based on topological insulator systems where the band gap has been tuned by substitution to a particular photonic band.   Our results show the intrinsic limitations on devices and their fabrication made for such purposes.

\section{methods}

The high quality epitaxial (Bi$_{1-x}$In$_x$)$_2$Se$_3$ thin-films were prepared by molecular beam epitaxy (MBE) using the two step growth recipe developed at Rutgers University \cite{Bansal11,Bansal12a}. An initial deposition of 3 QL was made at 110 $^{\circ}$C, which was followed by annealing to 220 
$^{\circ}$C, where the remaining deposition was performed to achieve the desired thickness. Reflection high energy electron diffraction (RHEED) was used to monitor the film quality during growth; RHEED indicates that the films are high quality single phase even when made as thin as 2 QL\cite{Bansal12a,Brahlek12}. The films were deposited in an excess Se flux, such that the flux ratio of Se to the combined Bi and In flux was kept $\sim$  10/1. Control of the Bi and In flux was critical to achieve the desired concentration and thickness. The Bi and In cells were calibrated $in$ $situ$ by a quartz crystal microbalance, and $ex$ $situ$ by Rutherford backscattering. Together these provide accuracy to within about $1\%$ of the targeted concentration. The precise substitution level can be confirmed as the low frequency infrared active phonon moves strongly to higher frequencies in a linear fashion with substitution of indium (SI Fig.4\textbf{f}). Samples were sealed in vacuum and sent immediately to JHU.

Standard TDTS in a transmission geometry was performed with a custom home-built THz spectrometer.   In this technique an almost single-cycle one picosecond pulse of electromagnetic radiation is transmitted through the sample.  The complex transmission function that results from Fourier transforming the pulse and ratioing it to a Fourier transformed reference pulse can be directly inverted to give the complex conductance $G(\omega)$ in the thin film limit: $\tilde{T}(\omega)=\frac{1+n}{1+n+Z_0G(\omega)} e^{i\Phi_s}$ where $\Phi_s$ is the phase accumulated from the small difference in thickness between the sample and reference substrates and $n$ is the substrate index of refraction.  By measuring both the magnitude and phase of the transmission, this inversion to conductance is done directly and does not require Kramers-Kronig transformation. TDTS turns out to be an ideal probe of the low frequency response of TI materials with both the metallic Drude transport peak and the
 phonon peak observable in the experimental range.  Low temperatures TDTS measurements generally began within 24 hours of their growth. The total exposure time to atmosphere was less than 30 minutes.  The samples were mounted inside an optical helium flow cryostat and cooled down to 5K within an hour. Transmission measurements in the mid-infrared (MIR) region were performed with a Fourier transform infrared spectrometer (Bruker Vertex 80) at room temperature in vacuum.

The data is fit by a Drude-Lorentz model by utilizing the \textit{RefFIT} program\cite{Kuzmenko_Reffit}. We include only a single Drude term (a Lorentzian centered at $\omega=0$), a Drude-Lorentz term (which models the phonon) and a frequency independent real $\epsilon_\infty$ contribution to the dielectric constant (that accounts for the effect of higher energy excitations on the low frequency physics)\cite{Rolando_Kerr, Rolando_Aging}: $G(\omega)=  \left(-\frac{\omega^{2}_{pD}}{i\omega-\Gamma_{D}}-\frac{i\omega\omega^{2}_{pDL}}{\omega^{2}_{DL}-\omega^{2}-i\omega\Gamma_{DL}}-i\left(\epsilon_{\infty}-1\right)\omega \right) \epsilon_0 d$ where $d$ is the film thickness. (See more discussion in SI section F).




\section{addendum}
We thank Y. Ando, N. Drichko, L. Fu, E.J. Mele, and F. Zhang for helpful discussions and F. Chen for help with AFM measurements.  The work at JHU was supported by the Gordon and Betty Moore Foundation through Grant GBMF2628 to NPA and from the DOE through DE-FG02-08ER46544.  The work at Rutgers was supported by IAMDN of Rutgers University, NSF DMR-0845464 and ONR N000140910749/N000141210456.

\textit{Competing Interests: }The authors declare that they have no
competing financial interests.

\textit{Correspondence: }Correspondence and requests for materials
should be addressed to NPA (email:npa@pha.jhu.edu).

\section{Author Contribution}

L.W performed the measurements and analyzed the data with the help and discussion from R.V.A, A.V.S, C.M.M and N.P.A; L.S.B and Y.L built the THz spectrometers. M.B, N.B and S.O synthesized the films. L.W and N.P.A wrote the manuscript from input of all authors. S.O and N.P.A devised the (Bi$_{1-x}$In$_x$)$_2$Se$_3$ experiment. L.W and N.P.A devised the ultrathin Bi$_2$Se$_3$ experiment.  All authors contributed to discussion on data analysis and manuscript editing.

\newpage
\begin{widetext}
\section{Supplementary Information for "A sudden collapse in the transport lifetime across the topological phase transition in (Bi$_{1-x}$In$_x$)$_2$Se$_3$``}

\bigskip

\subsection{Energy independence of evanescent depth}

As discussed in the main text, a surprising result of our analysis is that the evanescent decay length of the surface states into the bulk of the TI for the states at $E_F$ (350 meV above the Dirac point) is found to be approximately the same as the evanescent decay length of the states at the Dirac point (as determined by the thickness where ARPES on thin films sees a gap open at the Dirac point).   Although not typically emphasized, this is actually the theoretical expectation as it is a special feature of the Dirac model that the evanescent length is independent of binding energy  \cite{ZhangMele12, ZhangMelePrivateComm}.

It is most easily shown in a toy model of the linearized Dirac Hamiltonian.

\begin{equation}
H(k_z,k_{||}) = \left[\begin{array}{cc} \Delta & \hbar  (v_zk_z - i v_Fk_{||} )   \\ \hbar  (v_zk_z + i v_Fk_{||} ) & -\Delta \end{array}\right].
\end{equation}

Solving for the energy eigenvalues of this Hamiltonian gives the expression for the bulk valence and conduction bands

\begin{equation}
E =  \pm \sqrt{\Delta^2 +  (\hbar v_F k_{||})^2 +   (\hbar v_z k_{z})^2 }.
\end{equation}

One can see that within this model, the massless Dirac surface state dispersion $E  =  \hbar v_F  k_{||}$ asymptotically approaches this bulk band structure at high $E$ but always stays below it.

For any modes with energies below the band continuum the waves are evanescent with an inverse penetration depth.

\begin{equation}
k_z = \sqrt{E^2 - \Delta^2 -  (\hbar v_F k_{||})^2  } /   \hbar v_z .
\end{equation}

Substituting the dispersion for the massless Dirac dispersion $E  =  \hbar v_F  k_{||}$ this expression gives remarkably $k_z = i \Delta /  \hbar v_z $ for $all$ energies.

This result runs counter to one's usual intuition than the evanescent length will increase as a localized state's binding energy decreases (in this case as the TSS asymptotically approaches the bulk band edge).   In the present case, one can understand this result heuristically by realizing that the effective mass also increases as a function of energy and it does so as to exactly cancel the decreasing binding energy.  This result of an energy independent evanescent length holds within more realistic models as long as one considers a high symmetry surface of a material like Bi$_2$Se$_3$ \cite{ZhangMele12, ZhangMelePrivateComm}.  It is also evident upon close inspection of density functional theory calculations\cite{ZhangW10}.

\begin{figure*}
\begin{center}
\includegraphics[width=1\textwidth]{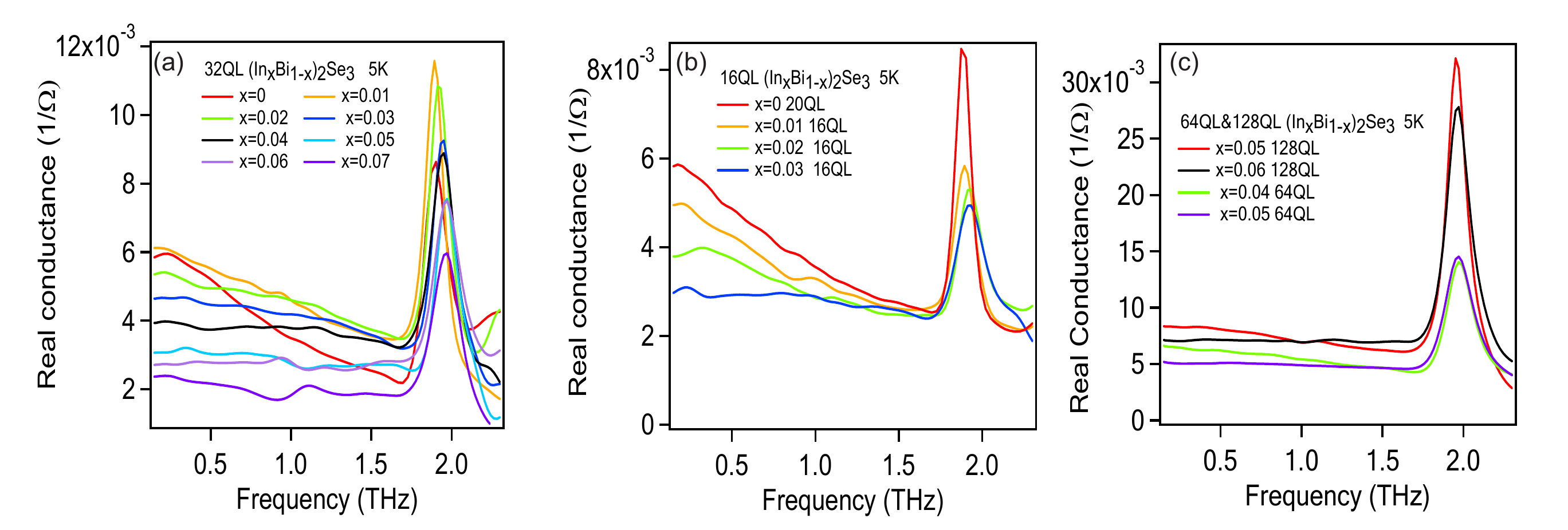}
\centering
\caption{Real parts of the THz sheet conductance of  \textsf{\textbf{a,}}  32QL  and  \textsf{\textbf{b,}}  16QL (Bi$_{1-x}$In$_x$)$_2$Se$_3$  films for different indium concentration levels at 5K.  \textsf{\textbf{c,}}  Real parts of the THz range sheet conductance for 128QL (Bi$_{1-x}$In$_x$)$_2$Se$_3$ with selective In concentration at 5K. 64QL samples are included for comparison to show finite size effect.}
\label{FinitesizeEff}
\end{center}
\end{figure*}

\begin{figure*}
\begin{center}
\includegraphics[width=0.8\textwidth]{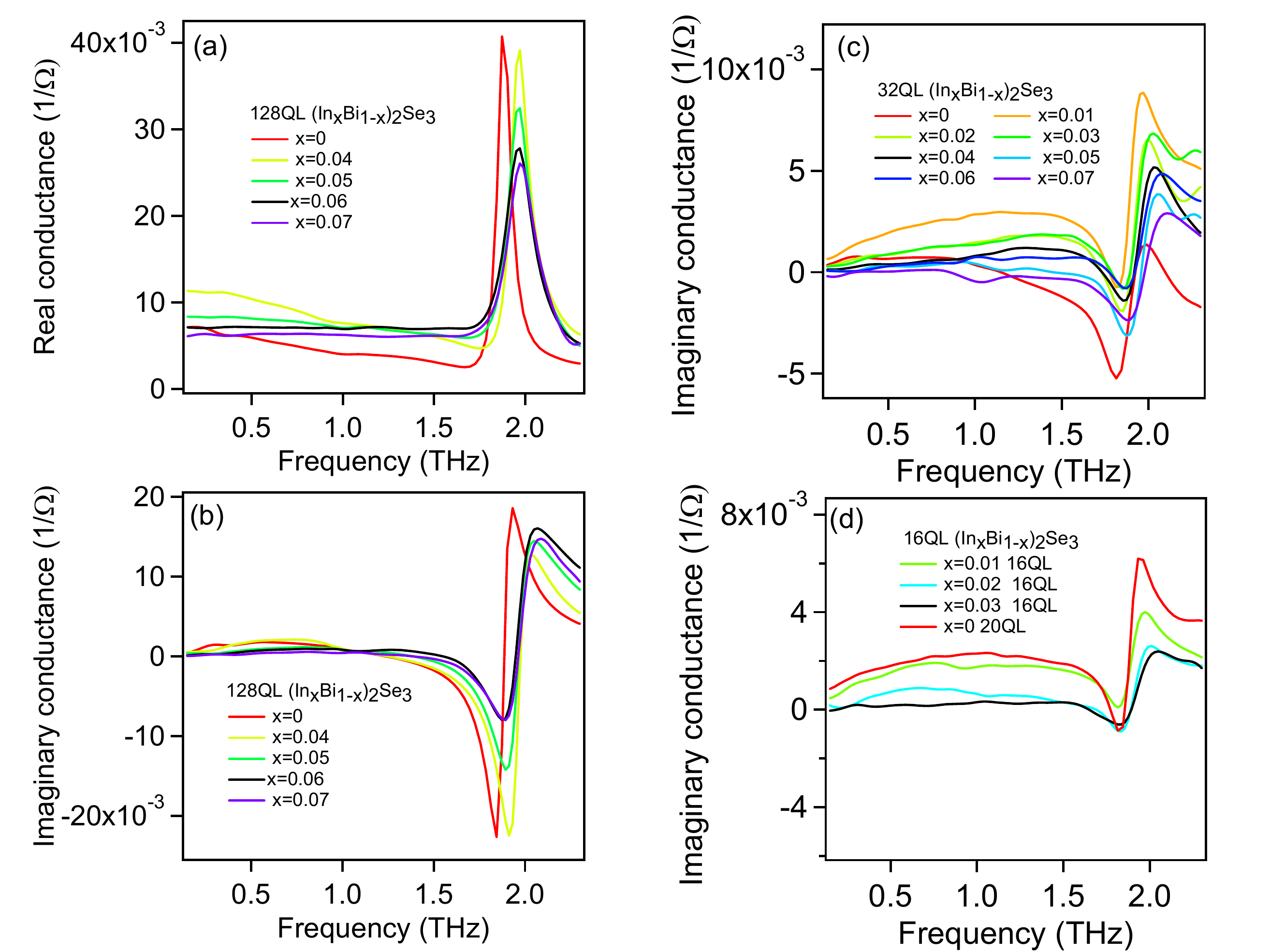}
\centering
\caption{  \textsf{\textbf{a,b,}}  Real and imaginary sheet conductance of 128QL (Bi$_{1-x}$In$_x$)$_2$Se$_3$ films at 5K.  \textsf{\textbf{c,d,}}  Imaginary sheet conductance of 32QL and 16QL  (Bi$_{1-x}$In$_x$)$_2$Se$_3$ films at 5K. }
\label{128QL32QL16QL}
\end{center}
\end{figure*}

\begin{figure*}
\begin{center}
\includegraphics[width=.8\textwidth]{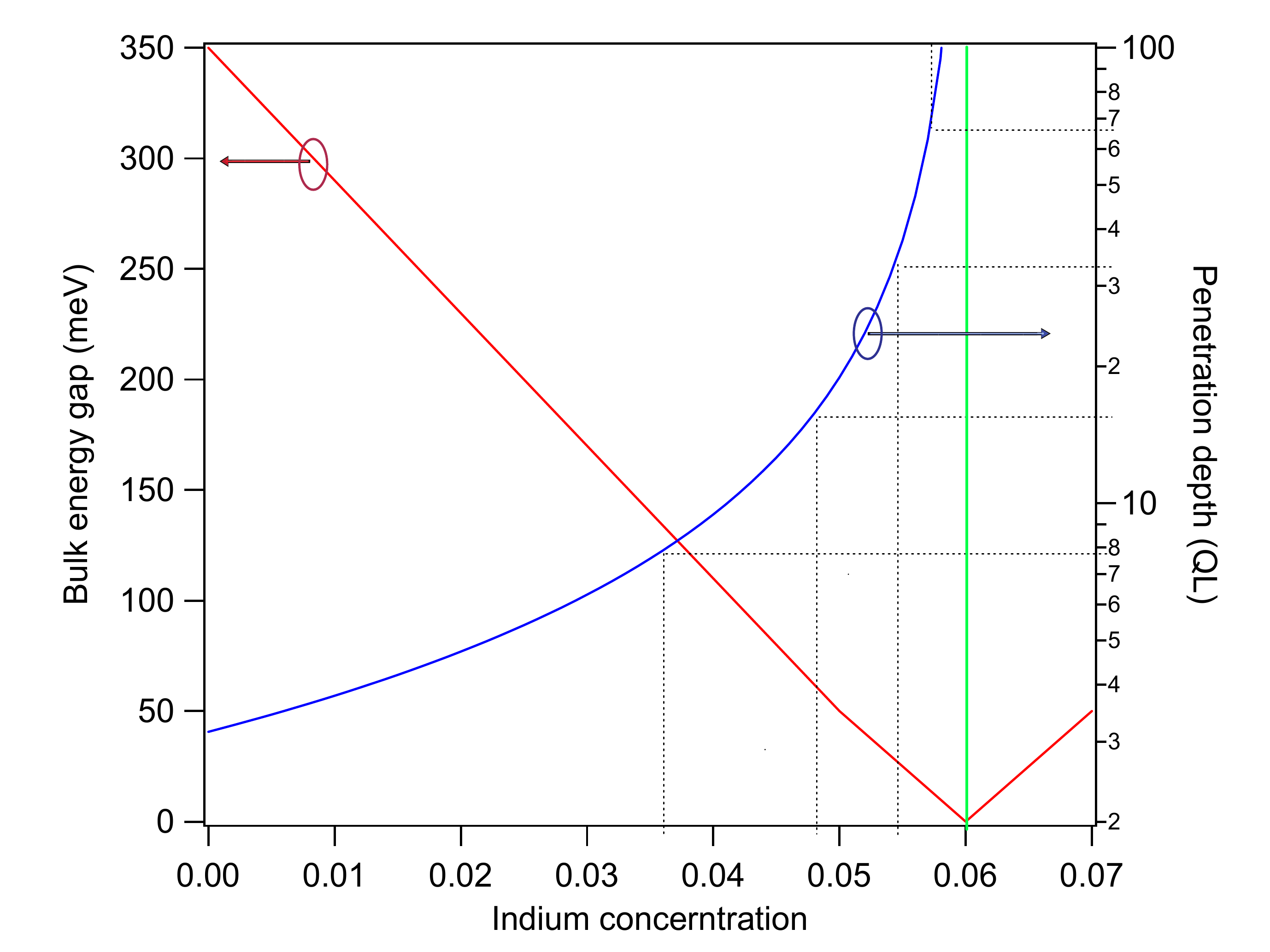}
\centering
\caption{Finite size analysis: surface state penetration depth (blue curve) as a function of In concentration based on linear extrapolation of the bulk energy gap (red curve) and the relation $\xi = \alpha \hbar v_F / \Delta$ where $\alpha$ is chosen to fit the penetration depth at $x=0$ inferred from Fig. 2\textsf{\textbf{b}} (main text).}
\label{Finitesize}
\end{center}
\end{figure*}

\subsection{Applicability of  (Bi$_{1-x}$In$_x$)$_2$Se$_3$ to investigate the topological phase transition to a conventional insulator }

As discussed in the main text of the paper, to investigate the quantum phase transition of a topological insulator to a trivial insulator it is interesting to investigate this physics in the most studied material class of Bi$_2$Se$_3$, especially in transport experiments.  In choosing a method to tune through this transition, the obvious substitution of the lighter element sulphur (S) onto the selenium (Se) site of the Bi$_2$Se$_3$ compound turns out to be inappropriate as Bi$_2$S$_3$ is orthorhombic with space group V$^{16}_h$ ($Pbnm$) with a twenty atom unit cell\cite{Larson02, Wyckoff86}.  This contrasts with the rhombohedral structure of  Bi$_2$Se$_3$ that has a symmetry
D$^{5}_{3d}$  ($R\overline{3}m$) with 5 atoms per unit cell.  This has led researchers to investigate the thallium-based
ternary chalcogenide alloy TlBi(S$_{1-x}$Se$_x$)$_2$ that maintains the same crystal structure throughout the  TlBiSe$_2$
$\rightarrow$  TlBiS$_2$  series.

It was pointed out recently that indium (In) substitutes for bismuth to form a solid solution in Bi$_2$Se$_3$ and that the non-topological end member In$_2$Se$_3$ of the (Bi$_{1-x}$In$_x$)$_2$Se$_3$ series shares the common rhombohedral  D$^{5}_{3d}$ structure with Bi$_2$Se$_3$\cite{Brahlek12}.  Indium substitution is particularly interesting as it is expected to have a large effect on the electronic properties of TIs because it substitutes for Bi, the heaviest element in Bi$_2$Se$_3$, and is therefore a very sensitive way to tune spin-orbit coupling.

\subsection{Finite size effects in (Bi$_{1-x}$In$_x$)$_2$Se$_3$ films}

In Fig. \ref{FinitesizeEff}\textsf{\textbf{a,b}}  we show the real part of THz sheet conductance for 32QL and 16QL (Bi$_{1-x}$In$_x$)$_2$Se$_3$ films. Recall that the low frequency conductance at 0.15 THz for 64QL (Bi$_{1-x}$In$_x$)$_2$Se$_3$ below $x=0.04$ differ very little from each other, but it systematically drops for  32QL (Bi$_{1-x}$In$_x$)$_2$Se$_3$ samples. Following curves with the same In substitution (curves with same color in Fig. \ref{FinitesizeEff}\textsf{\textbf{a,b}} for 32QL and 16QL (Bi$_{1-x}$In$_x$)$_2$Se$_3$, we can see the Drude component is more broadened for the 16QL samples. In Fig. \ref{FinitesizeEff}\textsf{\textbf{c}}, we show that the 64QL $x=0.04$ curve and the 128QL $x=0.05$ curve have the same width while the 64QL $x=0.05$ curve and the 128QL $x=0.06$ curve show the same flattening behavior.   This confirms the good quality of fit results shown in Fig. 2\textsf{\textbf{a}} (main text).  Considering the disorder level is a constant for the same In substitution, one can see the Drude peak is more broadened in thinner samples due to the finite size effect. The full set of data on 128QL (Bi$_{1-x}$In$_x$)$_2$Se$_3$ films and the imaginary part of the sheet conductance of 32QL and 16QL (Bi$_{1-x}$In$_x$)$_2$Se$_3$ films for different In concentrations at 5K are shown in Fig. \ref{128QL32QL16QL}. Note that the real part of the low frequency conductance of the $x=0.04$ 128QL film is a little higher than other films, which could come from slightly more bulk carriers in this film. As we will discuss in the below section \ref{SecFits}, we believe the bulk carriers give only a featureless background to the conductance within the measured THz spectral range and do not affect the fits appreciably.

For Fig.\ref{Finitesize}, we make a simple assumption that the bulk band gap linearly decreases from 350meV at $x=0$ to 0meV at $x=0.06$. The relation for surface state penetration  is $\xi = \alpha \hbar v_F / \Delta$ where $\alpha$ is a constant chosen to fit the penetration depth at $x=0$ inferred from Fig. 2\textsf{\textbf{b}}  (main text). The black dashed line indicates the concentration level when penetration depth equals to half of the film thickness for 16, 32, 64 and 128QL samples respectively.   The green line indicates the universal critical In concentration for band inversion, which is inferred from thickness independent MIR peak shown in Fig. 1\textsf{\textbf{b}}  (main text).  This simple picture based on a comparison of length scales accounts very well for the crossovers seen in Fig. 3\textsf{\textbf{f}}  (main text).

\subsection{The role of backscattering in thin TI films}

Previous calculations have shown the role that hybridization plays in opening a gap at the $\Gamma$ point when film thickness is reduced \cite{ZhangY10}.   Within a Fermi's golden rule approach this hybridization shows up in transport as an increased scattering probability, because isolated TSSs do not allow 180$^\circ$ backscattering.  However since top and bottom surfaces have opposite spin chiralities, hybridization opens up an expanded phase space for scattering when they spatially overlap. 

The increased phase space for scattering can be seen in a simple calculation.   The scattering rate determined by transport is generally less than the overall quasiparticle scattering rate, because transport depends on the momentum relaxation rate and hence is preferentially not sensitive to forward scattering.   In the simplest case of a circular Fermi surface and momentum independent impurity scattering, if the quasi-particle lifetime is $\tau_{qp}$, the elastic scattering rate measured in transport given within Fermi's golden rule is 

\begin{eqnarray}
1/\tau_{trans} = (1/\tau_{qp}) \frac{1}{2\pi} \int d \theta \Big( \frac {1 - \mathrm{cos} \theta}{2} \Big) \\
1/\tau_{trans} = \frac{1}{2}(1/\tau_{qp}) 
\end{eqnarray}

Due to the spin-chirality the transport scattering rate in a TSS is reduced by an additional factor that accounts for the overlap of spin states $|<S(k)|S(k+q)>|^2$ to be

\begin{eqnarray}
1/\tau_{trans}^{TSS} = (1/\tau_{qp}) \frac{1}{2\pi} \int d \theta \Big(\frac {1 - \mathrm{cos} \theta}{2} \Big) \Big (\frac {1 + \mathrm{cos} \theta}{2} \Big ) \\
1/\tau_{trans}^{TSS}  = \frac{1}{8}(1/\tau_{qp}) 
\end{eqnarray}

Therefore, the expectation within this very simple model is that when the wavefunctions from top and bottom surfaces start to overlap an electron can scatter into its momentum reversed state (See the inset of Fig. 2\textsf{\textbf{a}} (main text)) and the phase space for scattering quadruples.  Experimentally one sees about a 50\% increase.   That the experiment shows a smaller increase may be expected because of the presence of non-spin-momentum locked trivial states from the conduction band at $E_F$.  In such a situation some back scattering is already allowed and so the jump can not be as large as it would be if $E_F$ was in the bulk band gap. 

In this simplified picture we have ignored any changes to the Rashba SOC on each surface that enforces the spin-momentum locking. This can not be strictly realistic for the case for very thin films, because as the TSSs become delocalized between surfaces they experience the opposite sense of the Rashba SOC effect and hence a smaller average effect. Nevertheless, this effect is subdominant for ``small" wavefunction overlaps.   Also note that this calculation is for an ideal free-standing film.  In reality the doping levels of front and back surfaces should be a little different.   Moreover for films on a substrate, inversion symmetry is lost and there should be a Rashba SOC effect that comes from the substrate potential that leads to an additional spin splitting \cite{ZhangY10}.

Note that, in general, reducing the thickness of a metallic film will also increase the scattering rate because surface scattering increases.  However, considering that the exponential dependence fit for the $x=0$ sample gives a characteristic penetration depth of  approximately 2.5QL, which agrees with what ARPES observes \cite{ZhangY10}, this possibility that this is the principle effect can be excluded.

\begin{figure*}
\begin{center}
\includegraphics[width=1\textwidth]{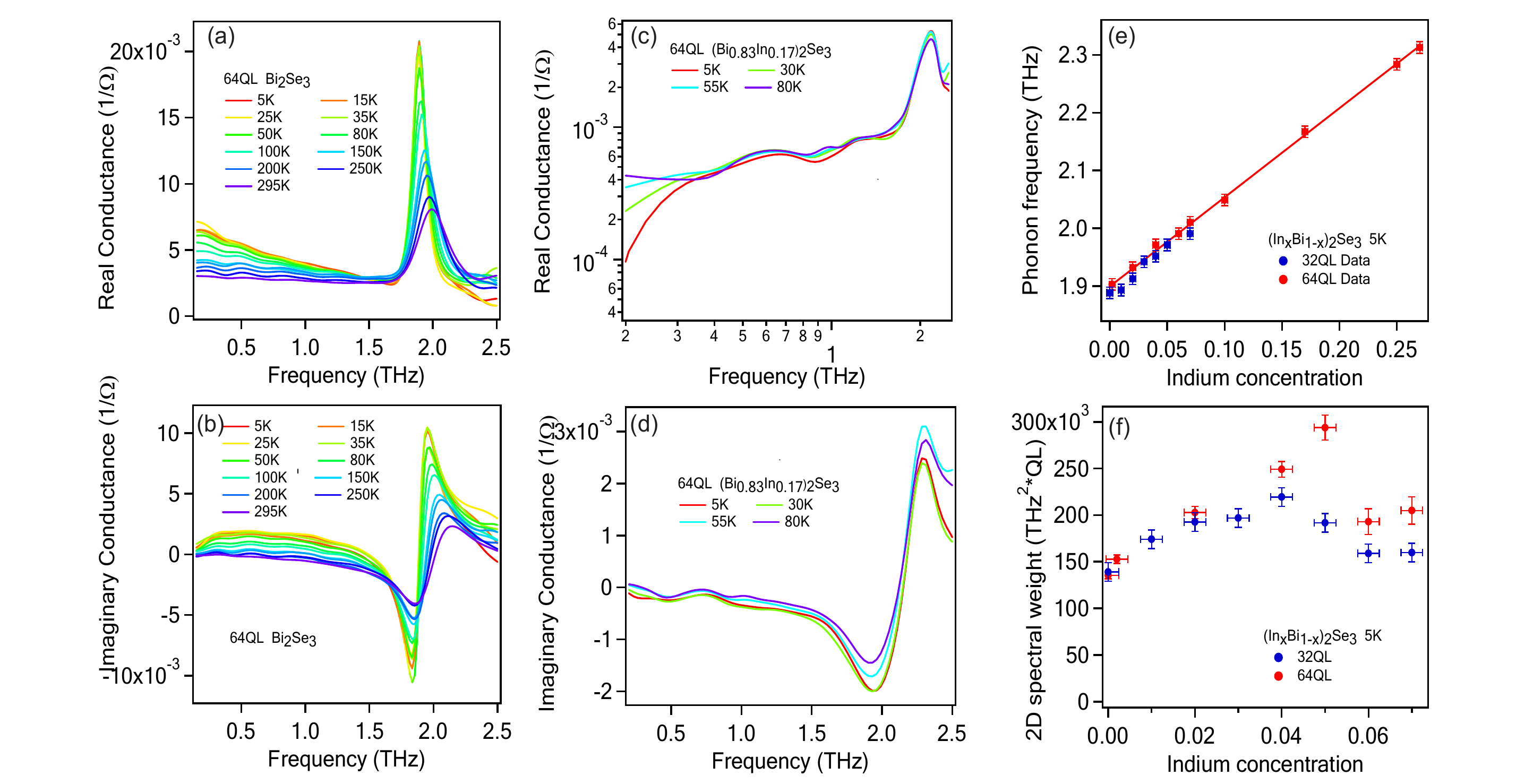}
\centering
\caption{ \textsf{\textbf{a,b,}}  Real and imaginary sheet conductance of Bi$_2$Se$_3$ films at different temperatures.  \textsf{\textbf{c,d,}} Real and imaginary sheet conductance of (Bi$_{0.83}$In$_0.17$)$_2$Se$_3$ films at different temperatures.  \textsf{\textbf{e,}}  Phonon frequency shift as a function of In substitution. The solid line is a linear fit to the 64QL data.   \textsf{\textbf{f,}}  2D Drude spectral weight ($(\omega_p/2\pi)^2 d$) as a function of In concentration for 32QL and 64QL (Bi$_{1-x}$In$_x$)$_2$Se$_3$ films.}
\label{TempDepen}
\end{center}
\end{figure*}

\subsection{THz conductivity of  (Bi$_{1-x}$In$_x$)$_2$Se$_3$ over a broader temperature and substitution range }

In Fig. \ref{TempDepen}\textsf{\textbf{a,b,}}  we show the temperature and frequency dependence of the complex conductance of pure Bi$_2$Se$_3$ film. The Drude peak broadens with increasing temperature, which is consistent with the expectation for conventional metallic transport. The phonon shifts slightly to higher frequency by 97 GHz ($\sim$3 cm$^{-1}$) from 5 K to room temperature. As noted earlier \cite{Laforge10}, such a large phonon hardening is unusual without a structural transition nearby in configuration space. Below 30K, the low frequency Drude peak has no notable temperature dependence and the Drude scattering rate is almost constant within our experimental accuracy. This, along with the fact that the DC resistance is constant below 30K\cite{Bansal12}, shows that even in the pure sample transport is dominated by impurity scattering. In contrast, (In$_{0.17}$Bi$_{0.83}$)$_2$Se$_3$ which is well into the topologically trivial regime shows increasing conductance as a function of frequency and the opposite temperature dependence at low frequency (Fig.  \ref{TempDepen}\textsf{\textbf{c}}) indicating the formation of a localized insulating state.  At this substitution, even the residual bulk conductance has disappeared, a metal-insulator transition by disorder induced localization has occurred, which agrees with DC transport data\cite{Brahlek12}.  In insulators with strong localization in the variable range hopping regime, the real part of the conductance typically shows a simple power-law behavior at low frequency $\omega$ \cite{Helgren04}, which interestingly is not exhibited in the measured spectral range in the present case. Fig.\ref{TempDepen}\textsf{\textbf{e}}  shows the phonon frequency linearly shifts with In concentration, which confirms the good quality of the sample.  Fig.\ref{TempDepen}\textsf{\textbf{f}} shows the spectral weight of the Drude component obtained from fits.

\begin{figure*}
\begin{center}
\includegraphics[width=1\textwidth]{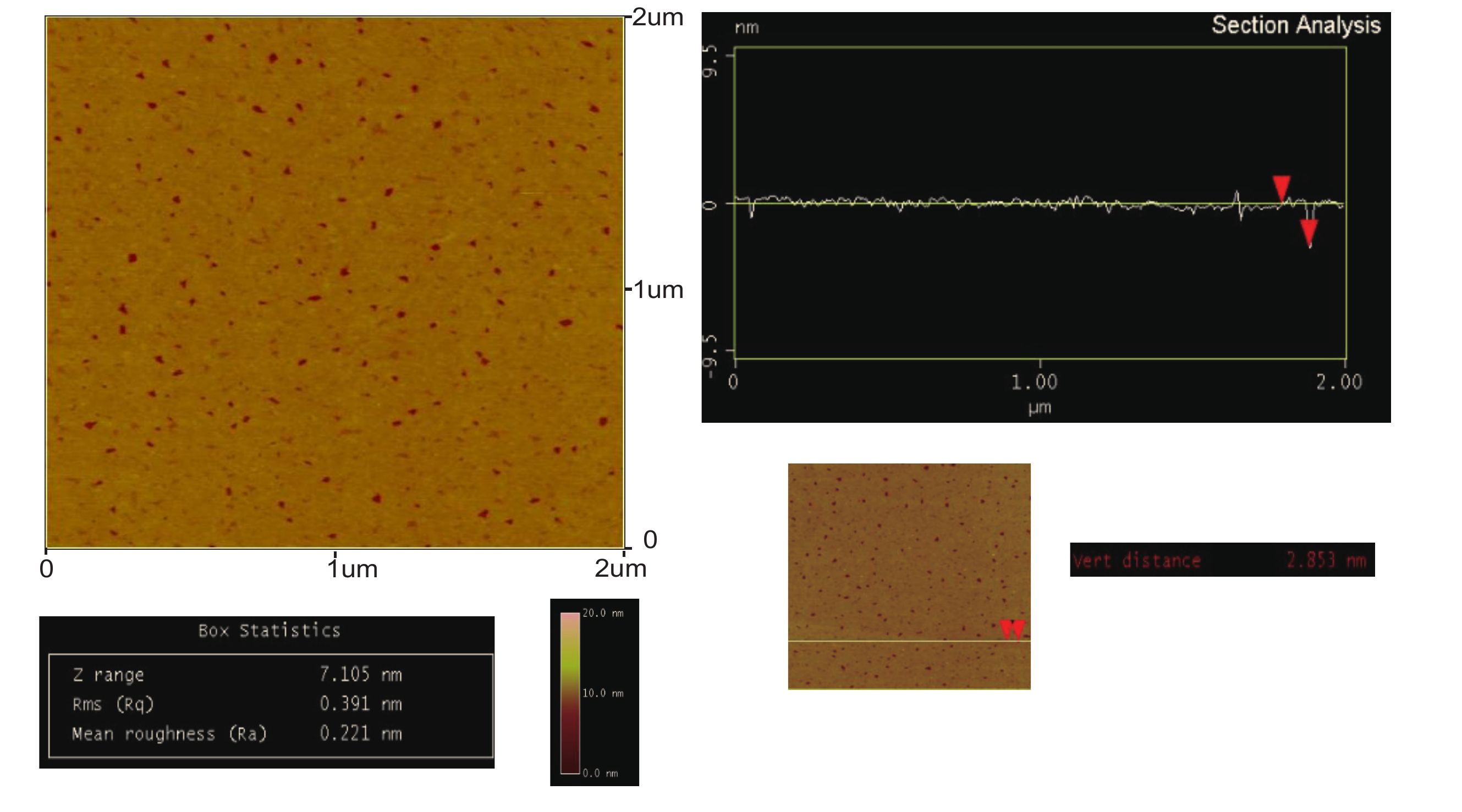}
\centering
\caption{AFM scan plot for a 2QL uncapped  Bi$_2$Se$_3$ sample. The statistics box shows the average roughness in the plot is 0.22nm and rms roughness is 0.39nm. On the right, we show the representative section analysis. The z-plot is taken along the white line indicated below. The biggest step is 2.853nm. We scanned different areas and they all show the similar results.}
\label{AFM}
\end{center}
\end{figure*}

\begin{figure*}
\begin{center}
\includegraphics[width=.8\textwidth]{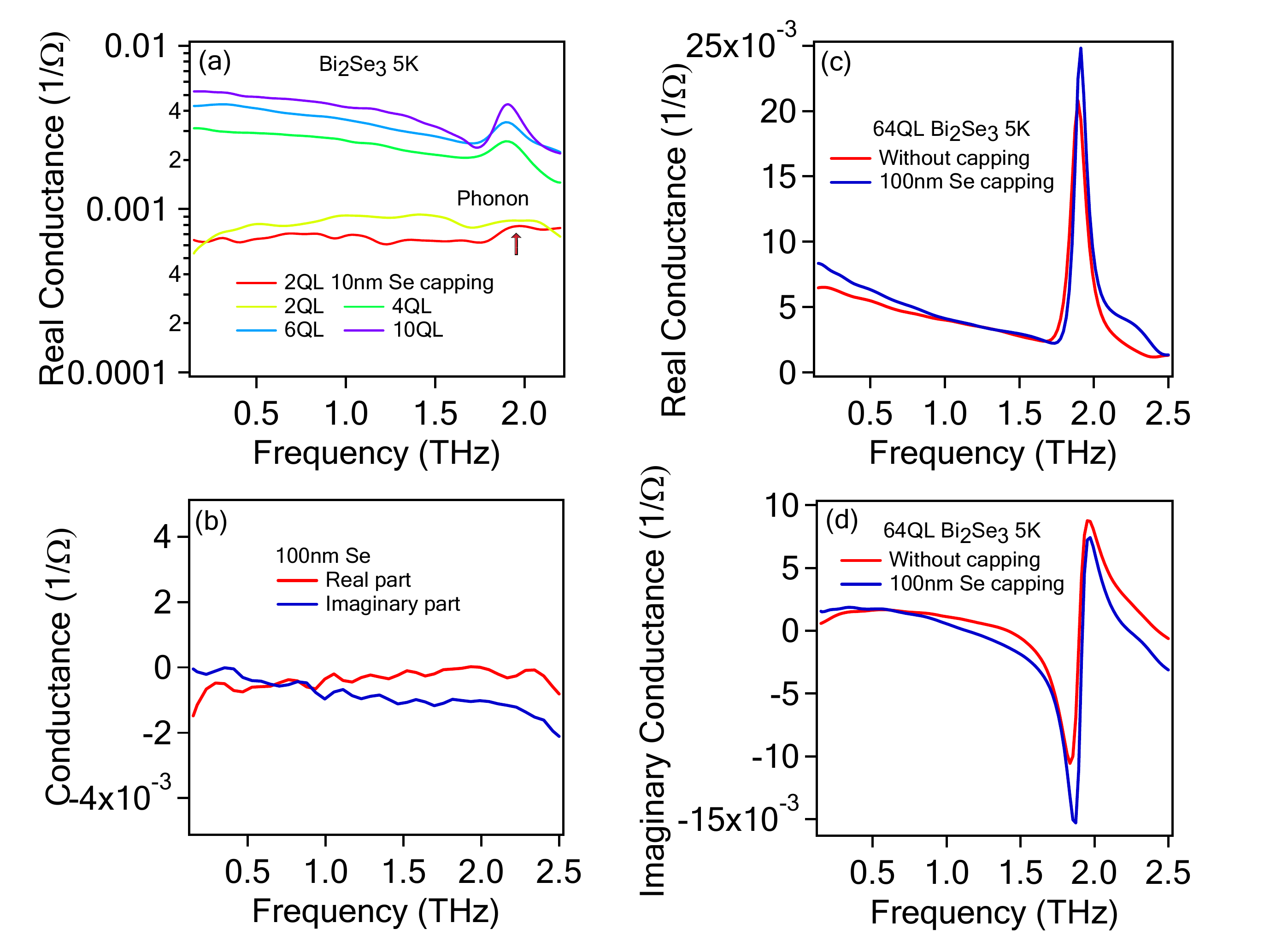}
\centering
\caption{ \textsf{\textbf{a,}} Real part of THz range sheet conductance for ultrathin Bi$_2$Se$_3$ samples at 5K.  \textsf{\textbf{b,}}  Real and imaginary THz sheet conductance for 100nm Se (grown on sapphire) at 5K  \textsf{\textbf{c,d,}}  Real and imaginary parts of THz sheet conductance of 64QL Bi$_2$Se$_3$  capped with 100nm Se and uncapped Bi$_2$Se$_3$ at 5K.}
\label{2QLSecapping}
\end{center}
\end{figure*}

\subsection{Note on Drude-Lorentz model fits }
\label{SecFits}
As shown by Fig. 1\textsf{\textbf{e}} (main text), the three component fits with a single Drude term is excellent and no other channels arising from bulk or surface accumulated regions are necessary.   The only possibilities to accommodate such channels are if these have scattering rates roughly coincident with the primary channel (which we consider unlikely), or more plausibly, if their scattering rates are so large as to make their contribution at low frequency small and featureless within the measured spectral range. This is what we observe in THz magneto-optics study on the Bi$_2$Se$_3$ film\cite{Wu_unpublished}.  In this study we found there are two Drude contributions: one channel we identify as TSSs channel (from the top and bottom surfaces combined) has a carrier density of $1.5 \times 10^{13}/cm^{2}$ (per surface) and scattering rate $1/\tau$ $\sim$ 1 THz and a second channel which may either come from a bulk or accumulation layer that has $7.5 \times 10^{12}/cm^{2}$ and scattering rate $1/\tau$ $\sim$ 9 THz.  The conductance ratio at zero frequency between the TSSs channel and the other channel is around 10. The scattering rate in the non-TI regime is still less than the value of the second channel. Therefore, one Drude component approximation is appropriate.

One might be concerned that the scattering rate of the principle Drude component is not very accurately determined when the scattering rate becomes large so that the real part of the conductance is flat within our spectral range. However, note that within the Drude model that when the real part of the conductance is flat, the imaginary part goes like $\omega \tau$ and so has a spectral dependence at low frequency. We are sensitive to the scattering rate for some limited range around $6\%$ due to this dependence.  The uncertainty in the scattering rate in this substitution range does increase however and is reflected in the fact that the errors bars in the plotted data increase in Fig. 2\textsf{\textbf{a}}  (main text).    However, the sharp increase of the scattering at the threshold is well above the uncertainty.    Note that our ultimate conclusions about the existence of the sharp threshold of an increase in scattering are independent of the precise value of the scattering rate above this threshold.   The important point is that we can identify the threshold.

We can estimate the Fermi energy and carrier density from the Drude component fit.   In optical conductivity measurements the integrated spectral weight of a feature in the real part of the conductance ($G_1=\sigma d$) is a function of the areal charge density $n_{2D}$ over the effective mass $m^{*}$.

\begin{equation}
\frac{2}{\pi\epsilon_{0}}\int G_1 d \omega = \omega_{p}^{2} d =\frac{n_{2D}e^{2}}{m^{*}\epsilon_{0}} 
\end{equation}

Here $d$ is the film thickness.  In our work we get the integrated spectral weight through its dependence of the fitting parameter $\omega_{p}$ in our Drude fits (See Methods).  The effective mass is expressed as $m^{*}=\hbar k_{F}/v_{F}$ and the Fermi velocity is determined by $v_{F}=\partial E_{F}/\hbar\partial k_{F}$.  Although the leading order dispersion of the TSSs is linear, an appreciable quadratic correction has been found to exist ($E = A k_F + B k_F^2$).  In Ref. \cite{Jenkins10}, the authors fit the dispersion of TSSs from Ref. \cite{XiaY09} and give an expression for the dispersion with quadratic corrections as $E_{F}=2.02 k_{F}+10.44 k_{F}^{2}$. Here $k_{F}$ is in unit of $\AA^{-1}$ and the coefficients have the units to give $E_{F}$ in eV.  For one surface state, $n_{2D}=k_{F}^{2}/4\pi$.  If we make the approximation that the two surfaces have equal densities and both contribute to the conductance equally the expression follows

\begin{equation}
\omega_{p}^{2} d  =  \frac{k_F ( A + 2B k_F)e^2 }{ 2 \pi \hbar^2 \epsilon_0}
\end{equation}

From this expression and using the data from our 64 QL $x=0$ Bi$_2$Se$_3$ sample, we get $k_{F}=0.15\AA^{-1}$,  the density $n=1.8 \times 10^{13}/cm^{2}$ (per surface), and an $E_{F}\sim 0.5eV$ above the Dirac point.   This $E_F$ is very close to the $E_F\sim0.45eV$ extracted from ARPES data on Bi$_2$Se$_3$ aged in UHV\cite{Bianchi10} and aged in air\cite{Analytis10}.

\subsection{Ultrathin 2QL Bi$_2$Se$_3$ sample}

As pointed out in the main text, a very large qualitative and quantitative difference is seen in the thickness dependence between 2 and 4QL films.  The possibility exists that  this reflects the predicted oscillatory behavior between 2D topologically non-trivial and trivial states as a function of thickness\cite{Lu-ShenPRB10, Liu-ZhangPRB10}.  However, it is reasonable to have concerns that 1 QL step edges or surface oxidation of such a thin sample may be the source of such a large difference.

The sample does not appear to have appreciable surface density of 1 QL step edges.  As mentioned in the methods section, the RHEED pattern is sharp even for the 2QL sample\cite{Bansal11, Bansal12}.  As shown in Fig.\ref{AFM}, the representative contour plot by AFM shows the average roughness is 0.221nm and rms roughness is 0.391nm and, therefore, the overall surface is reasonably flat. The representative section analysis further confirms the flatness of surface over a wide area. Therefore, the surface roughness is not a significant concern in this sample.

To alleviate concerns about surface oxidation, we also measured a 2QL  Bi$_2$Se$_3$ film capped with 10nm Se as a test sample. An additional concern is that although Se capping can protect the surface from oxidization, one might worry that amorphous Se  could contribute an additional conducting channel. To test for free carriers in amorphous Se, we measured a 100nm ($\pm 1$ nm) Se sample grown on sapphire by MBE.  Fig.\ref{2QLSecapping}\textsf{\textbf{b}} shows that the real conductance is almost zero and the imaginary part has a slight negative slope, which together shows insulating behavior of the Se layer. Therefore, a Se capping layer is good for preventing oxidization without contributing an additional conducting channel. Note that the real conductance is negative at some frequency regime because the very low conductance of insulating Se layer is close to our detection limit.

To show the essentially inert character of the Se capping layer, we measured a 64QL Bi$_2$Se$_3$ film capped with 100nm Se. Fig.\ref{2QLSecapping}\textsf{\textbf{c,d}} shows that the difference between uncapped and capped Bi$_2$Se$_3$ films is small. This shows the protective ability of the Se capping layer.  Since Se protects the film well, we can conclude that the principal difference between 4 and 2 QL films is likely not from surface oxidation.   Fig.\ref{2QLSecapping}\textsf{\textbf{a}}  shows the data on another 2QL Bi$_2$Se$_3$ capped with 10nm Se. We see the phonon peak with the same frequency of thicker films, which confirms the good quality of structure in the 2QL sample. Similar behavior is also seen in the 64QL Bi$_2$Se$_3$ sample capped with 100nm Se. Because the conductance of the 2QL sample (both capped and uncapped) is close to the low limit we can detect, the signal-to-noise ratio is not good enough and we can not perform reliable fits on it to extract the Drude spectral weight and scattering rate, etc. However, the phonon peak is unmistakably apparent.

To summarize, we believe the sample quality of 2QL film is reasonably good. We tried to rule out the possibility of surface oxidization and have demonstrated the flatness of surface.   The large difference between 2QL and 4 QL samples is as of yet unexplained.  Even though we cannot extract reliable parameters from fits, considering the huge difference between 2QL and 4QL samples, a possible scenario is the predicted oscillatory behavior between 2D topologically non-trivial and trivial states as a function of thickness\cite{Lu-ShenPRB10, Liu-ZhangPRB10}.



\newpage
\end{widetext}

\end{document}